\documentclass{article}
\usepackage{amsmath}

 \evensidemargin0in \oddsidemargin0in \topmargin10pt
\begin{document}

\title{Collins diffraction formula and the Wigner function in entangled
state representation }
\author{Hong-Yi Fan$^{1,2}$ and Li-yun Hu$^{1\text{*}}$ \\
%EndAName
$^{1}${\small Department of Physics, Shanghai Jiao Tong University, Shanghai
200030, China }\\
$^{2}${\small Department of Material Science and Engineering, }\\
{\small University of Science and Technology of China, Hefei, Anhui 230026,
China}\\
{\small Corresponding author. E-mail address: hlyun2008@126.com.}}
\maketitle

\begin{abstract}
{\small Based on the correspondence between Collins diffraction formula
(optical Fresnel transform) and the transformation matrix element of a
three-parameters two-mode squeezing operator in the entangled state
representation (Opt. Lett. 31 (2006) 2622) we further explore the
relationship between output field intensity determined by the Collins
formula and the input field's probability distribution along an infinitely
thin phase space strip both in spacial domain and frequency domain. The
entangled Wigner function is introduced for recapitulating the result.}

\end{abstract}

In a preceding Letter \cite{1} we have reported that the Collins diffraction
formula in cylindrical coordinates is just the transformation matrix element
of a three-parameter ($k$ and $t$ are complex and satisfy the unimodularity
condition $kk^{\ast }-tt^{\ast }=1)$ two-mode squeezing operator \cite{2,3}%
\begin{equation}
F^{\left( t,k\right) }=\exp \left( \frac{t}{k^{\ast }}a_{1}^{\dagger
}a_{2}^{\dagger }\right) \exp \left[ \left( a_{1}^{\dagger
}a_{1}+a_{2}^{\dagger }a_{2}+1\right) \ln \left( k^{\ast }\right) ^{-1}%
\right] \exp \left( -\frac{t^{\ast }}{k^{\ast }}a_{1}a_{2}\right) ,
\label{1}
\end{equation}%
in the deduced entangled state representation $\left\langle s,r^{\prime
}\right\vert ,$
\begin{eqnarray}
\phi _{s}\left( r^{\prime }\right)  &\equiv &\left\langle s,r^{\prime
}\right\vert \left. \phi \right\rangle =\left\langle s,r^{\prime
}\right\vert F^{\left( t,k\right) }\left\vert \psi \right\rangle   \notag \\
&=&\frac{\mathtt{i}^{s}}{2\mathtt{i}B}\int_{0}^{\infty }\mathtt{d}\left(
r^{2}\right) \exp \left[ \frac{\mathtt{i}}{2B}\left( Ar^{2}+Dr^{\prime
2}\right) \right] J_{s}\left( -\frac{rr^{\prime }}{B}\right) \psi _{s}\left(
r\right) ,  \label{2}
\end{eqnarray}%
where $\psi _{s}$ and $\phi _{s}$ denote the incoming and output light,
respectively, $[a_{i},a_{j}^{\dagger }]=\delta _{i,j},$%
\begin{equation}
k=\frac{1}{2}\left[ A+D-\mathtt{i}\left( B-C\right) \right] ,\text{\ \ }t=%
\frac{1}{2}\left[ A-D+\mathtt{i}\left( B+C\right) \right] ,  \label{3}
\end{equation}%
we see that the relation $kk^{\ast }-tt^{\ast }=1$ becomes $AD-BC=1$, $J_{s}$
is the $s$th Bessel function, and
\begin{equation}
\left\langle s,r^{\prime }\right\vert =\frac{1}{2\pi }\int_{0}^{2\pi }%
\mathtt{d}\theta e^{\mathtt{i}s\theta }\left\langle \eta =r^{\prime }e^{%
\mathtt{i}\theta }\right\vert ,  \label{4}
\end{equation}%
here $\left\vert \eta \right\rangle $ is the entangled states in two-mode
Fock space \cite{4,5,6,7} named after Einstein-Podolsky-Rosen (EPR)'s \cite%
{8} concept of quantum entanglement,
\begin{equation}
\left\vert \eta \right\rangle =\exp \{-\frac{1}{2}\left\vert \eta
\right\vert ^{2}+\eta a_{1}^{\dagger }-\eta ^{\ast }a_{2}^{\dagger
}+a_{1}^{\dagger }a_{2}^{\dagger }\}\left\vert 00\right\rangle .  \label{5}
\end{equation}%
Thus $\left\langle s,r^{\prime }\right\vert F^{\left( t,k\right) }\left\vert
\psi \right\rangle $ is the quantum optics version of the Collins formula
(generalized Hankel transformation). In \cite{9} we have also found
\begin{equation}
\mathcal{K}^{\left( t,k\right) }\left( \eta ^{\prime },\eta \right) =\frac{1%
}{\pi }\left\langle \eta ^{\prime }\right\vert F^{\left( t,k\right)
}\left\vert \eta \right\rangle =\frac{1}{2\mathtt{i}B\pi }\exp \left\{ \frac{%
\mathtt{i}}{2B}\left[ A\left\vert \eta \right\vert ^{2}-\left( \eta \eta
^{\prime \ast }+\eta ^{\ast }\eta ^{\prime }\right) +D\left\vert \eta
^{\prime }\right\vert ^{2}\right] \right\} .  \label{6}
\end{equation}%
Comparing with the integral kernel of usual Fresnel transform which
describes how a general beam $\psi \left( x^{\prime }\right) ,$ propagating
through an $\left( ABCD\right) $ optical paraxial system, becomes output
field $\phi \left( x\right) $\cite{10,11}%
\begin{equation}
\phi \left( x\right) =\int_{-\infty }^{\infty }\mathcal{K}\left( x,x^{\prime
}\right) \psi \left( x^{\prime }\right) \mathtt{d}x^{\prime },  \label{8}
\end{equation}%
where $AD-BC=1,$%
\begin{equation}
\mathcal{K}\left( x,x^{\prime }\right) =\frac{1}{\sqrt{2\pi \mathtt{i}B}}%
\exp \left[ \frac{\mathtt{i}}{2B}\left( Ax^{\prime 2}-2x^{\prime
}x+Dx^{2}\right) \right] ,  \label{9}
\end{equation}%
we see that $\mathcal{K}^{\left( t,k\right) }\left( \eta ^{\prime },\eta
\right) $ can be considered as the integration kernel of 2-dimensional
entangled optical Fresnel transform,%
\begin{equation}
\Psi \left( \eta ^{\prime }\right) =\int \mathcal{K}^{\left( t,k\right)
}\left( \eta ^{\prime },\eta \right) \Phi \left( \eta \right) \mathtt{d}%
^{2}\eta ,  \label{7}
\end{equation}%
in this sense $F^{\left( t,k\right) }$ can be named entangled Fresnel
operator (EFO), here $\Phi \left( \eta \right) =\left\langle \eta
\right\vert \left. \Phi \right\rangle ,$ $\Psi \left( \eta ^{\prime }\right)
=\left\langle \eta ^{\prime }\right\vert \left. \Psi \right\rangle ,$ and we
have used the completeness relation $\int \frac{\mathtt{d}^{2}\eta }{\pi }%
\left\vert \eta \right\rangle \left\langle \eta \right\vert =1.$

Clearly, if the $\left[ ABCD\right] $ system is changed to $\left[ D\left(
-B\right) \left( -C\right) A\right] $ system, then Eq. (\ref{7}) should read%
\begin{equation}
\Psi \left( \eta ^{\prime }\right) =\int \mathcal{K}_{2}^{\left(
D,-B,-C\right) }\left( \eta ^{\prime },\eta \right) \Phi \left( \eta \right)
\mathtt{d}^{2}\eta ,  \label{10}
\end{equation}%
where $\mathcal{K}_{2}^{\left( D,-B,-C\right) }$ is%
\begin{equation}
\mathcal{K}_{2}^{\left( D,-B,-C\right) }\left( \eta ^{\prime },\eta \right) =%
\frac{1}{-2\mathtt{i}B\pi }\exp \left\{ \frac{\mathtt{i}}{-2B}\left[
D\left\vert \eta \right\vert ^{2}-\left( \eta \eta ^{\prime \ast }+\eta
^{\ast }\eta ^{\prime }\right) +A\left\vert \eta ^{\prime }\right\vert ^{2}%
\right] \right\} .  \label{11}
\end{equation}

On the other hand, signals or images in optical information theory may be
described directly or indirectly by the Wigner distribution function (WDF)
\cite{12}. In one-dimensional (1D) case, the WDF of an optical signal field $%
\psi \left( x\right) $ is defined as
\begin{equation}
W_{\psi }(\nu ,x)=\int_{-\infty }^{+\infty }\frac{\mathtt{d}u}{2\pi }e^{%
\mathtt{i}\nu u}\psi ^{\ast }\left( x+\frac{u}{2}\right) \psi \left( x-\frac{%
u}{2}\right) .  \label{12}
\end{equation}%
$W_{\psi }(\nu ,x)$ involves both spatial distribution information and
space-frequency distribution information of the signal, $\nu $ is named
space frequency. Now, let us consider the entangled case. Like Eq. (\ref{12}%
), it is natural to introduce the 2-D complex Wigner transform as
\begin{equation}
W\left( \sigma ,\gamma \right) =\int \frac{\mathtt{d}^{2}\eta }{\pi ^{3}}%
\psi \left( \sigma +\eta \right) \psi ^{\ast }\left( \sigma -\eta \right)
e^{\eta \gamma ^{\ast }-\eta ^{\ast }\gamma },  \label{13}
\end{equation}%
where $\sigma ,$ $\gamma ,$ $\eta $ are all complex variables. To see its
physical meaning, using the integration formula of Dirac $\delta -$function,
we perform the following integration,
\begin{equation}
\int \mathtt{d}^{2}\gamma W\left( \sigma ,\gamma \right) =\int \frac{\mathtt{%
d}^{2}\eta }{\pi }\psi \left( \sigma +\eta \right) \psi ^{\ast }\left(
\sigma -\eta \right) \delta \left( \eta \right) \delta \left( \eta ^{\ast
}\right) =\frac{1}{\pi }\left\vert \psi \left( \sigma \right) \right\vert
^{2},  \label{14}
\end{equation}%
which is just the probability distribution of the complex function $\psi $($%
\sigma $). Further, let the ordinary Fourier transforms of $\psi \left(
\sigma \right) $ be $j\left( \zeta \right) ,$%
\begin{equation}
\psi \left( \sigma \right) =\int \frac{\mathtt{d}^{2}\zeta }{2\pi }j\left(
-\zeta \right) e^{\left( \zeta ^{\ast }\sigma -\zeta \sigma ^{\ast }\right)
/2},  \label{15}
\end{equation}%
then substituting (\ref{15}) into (\ref{13}) leads to%
\begin{eqnarray}
W\left( \sigma ,\gamma \right)  &=&\int \frac{\mathtt{d}^{2}\eta }{\pi ^{3}}%
\frac{\mathtt{d}^{2}\zeta }{2\pi }\frac{\mathtt{d}^{2}\zeta ^{^{\prime }}}{%
2\pi }j\left( -\zeta \right) j^{\ast }\left( -\zeta ^{^{\prime }}\right) e^{%
\frac{\left( \zeta ^{\ast }-\zeta ^{/\ast }\right) \sigma -\left( \zeta
-\zeta ^{/}\right) \sigma ^{\ast }}{2}}e^{\eta \left( \gamma ^{\ast }+\frac{%
\zeta ^{\ast }+\zeta ^{/\ast }}{2}\right) -\eta ^{\ast }\left( \gamma +\frac{%
\zeta +\zeta ^{^{\prime }}}{2}\right) }  \notag \\
&=&\int \frac{\mathtt{d}^{2}\zeta }{\pi ^{3}}j\left( -\zeta \right) j^{\ast
}\left( 2\gamma +\zeta \right) e^{\left( \zeta ^{\ast }+\gamma ^{\ast
}\right) \sigma -\left( \zeta +\gamma \right) \sigma ^{\ast }}=\int \frac{%
\mathtt{d}^{2}\zeta }{\pi ^{3}}j\left( \gamma -\zeta \right) j^{\ast }\left(
\gamma +\zeta \right) e^{\zeta ^{\ast }\sigma -\zeta \sigma ^{\ast }}.
\label{16}
\end{eqnarray}%
It then follows from (\ref{16}) that%
\begin{eqnarray}
\int \mathtt{d}^{2}\sigma W\left( \sigma ,\gamma \right)  &=&\int \frac{%
\mathtt{d}^{2}\zeta }{\pi ^{3}}j\left( \gamma -\zeta \right) j^{\ast }\left(
\zeta +\gamma \right) \int \mathtt{d}^{2}\sigma e^{\zeta ^{\ast }\sigma
-\zeta \sigma ^{\ast }}  \notag \\
&=&\int \frac{\mathtt{d}^{2}\zeta }{\pi }j\left( \gamma -\zeta \right)
j^{\ast }\left( \zeta +\gamma \right) \delta \left( \zeta \right) \delta
\left( \zeta ^{\ast }\right) =\frac{1}{\pi }\left\vert j\left( \gamma
\right) \right\vert ^{2},  \label{17}
\end{eqnarray}%
which is the probability distribution of the complex function $j\left(
\gamma \right) $. Thus our definition in (\ref{13}) leads to two marginal
distributions in $\sigma $ and $\gamma $ phase space, respectively. Hence $%
W\left( \sigma ,\gamma \right) $ is indeed the correct complex 2-D Wigner
function (Wigner transform) of complex function $\psi \left( \sigma \right) $
or $j\left( \gamma \right) $. If one wants to reconstruct the Wigner
function by using various probability distributions, obviously the
\textquotedblleft position density" $\left\vert \left\langle \sigma
\right\vert \left. \psi \right\rangle \right\vert ^{2}$ and the
space-frequency density $\left\vert \left\langle \gamma \right\vert \left.
\psi \right\rangle \right\vert ^{2}$ are not enough, so we extend $\delta
\left( \eta \right) \delta \left( \eta ^{\ast }\right) \equiv \delta \left(
\eta _{1}\right) \delta \left( \eta _{2}\right) $ to $\delta \left( \eta
_{1}-D\sigma _{1}-B\gamma _{2}\right) \delta \left( \eta _{2}-D\sigma
_{2}+B\gamma _{1}\right) $ and generalize (\ref{14}) to$,$%
\begin{equation}
R_{2}\left( \eta _{1},\eta _{2}\right) \equiv \pi \int \delta \left( \eta
_{1}-D\sigma _{1}-B\gamma _{2}\right) \delta \left( \eta _{2}-D\sigma
_{2}+B\gamma _{1}\right) W\left( \sigma ,\gamma \right) \mathtt{d}^{2}\sigma
\mathtt{d}^{2}\gamma ,  \label{18}
\end{equation}%
$R_{2}\left( \eta _{1},\eta _{2}\right) $ is also a probability distribution
along an infinitely thin phase space strip denoted by the real parameters $%
B,D$, which is a generalized entangled Radon transform \cite{13,14} of the
two-mode Wigner function (in the entangled form) \cite{15,16},

Then an interesting question naturally arises: what is the relation between
the generalized Fresnel transform and the WDF in entangled state
representation?

We begin with rewriting the 2-D WF (\ref{13}) as
\begin{eqnarray}
W\left( \sigma ,\gamma \right)  &=&\int \mathtt{d}^{2}\sigma ^{\prime }%
\mathtt{d}^{2}\sigma ^{\prime \prime }\int \frac{\mathtt{d}^{2}\eta }{\pi
^{3}}\psi \left( \sigma ^{\prime }\right) \psi ^{\ast }\left( \sigma
^{\prime \prime }\right) \delta ^{(2)}\left( \sigma ^{\prime }-\sigma -\eta
\right) \delta ^{(2)}\left( \sigma -\eta -\sigma ^{\prime \prime }\right)
e^{\eta \gamma ^{\ast }-\eta ^{\ast }\gamma }  \notag \\
&=&\int \frac{\mathtt{d}^{2}\sigma ^{\prime }\mathtt{d}^{2}\sigma ^{\prime
\prime }}{\pi ^{3}}\psi \left( \sigma ^{\prime }\right) \psi ^{\ast }\left(
\sigma ^{\prime \prime }\right) \delta ^{\left( 2\right) }\left( 2\sigma
-\sigma ^{\prime }-\sigma ^{\prime \prime }\right) e^{\left( \sigma ^{\prime
}-\sigma \right) \gamma ^{\ast }-\left( \sigma ^{\prime }-\sigma \right)
^{\ast }\gamma }.  \label{19}
\end{eqnarray}%
Substituting (\ref{19}) into (\ref{18}) we rewrite the Radon transform of $%
W\left( \sigma ,\gamma \right) $ as ($\mathtt{d}^{2}\sigma =\mathtt{d}\sigma
_{1}\mathtt{d}\sigma _{2},$ $\mathtt{d}^{2}\gamma =\mathtt{d}\gamma _{1}%
\mathtt{d}\gamma _{2}$)%
\begin{eqnarray}
R_{2}\left( \eta _{1},\eta _{2}\right)  &=&\int \frac{\mathtt{d}^{2}\sigma
^{\prime }\mathtt{d}^{2}\sigma ^{\prime \prime }}{\pi ^{2}}\psi \left(
\sigma ^{\prime }\right) \psi ^{\ast }\left( \sigma ^{\prime \prime }\right)
\int \mathtt{d}^{2}\sigma \mathtt{d}^{2}\gamma \delta \left( \eta
_{2}-D\sigma _{2}+B\gamma _{1}\right)   \notag \\
&&\times \delta \left( \eta _{1}-D\sigma _{1}-B\gamma _{2}\right) \delta
^{\left( 2\right) }\left( 2\sigma -\sigma ^{\prime }-\sigma ^{\prime \prime
}\right) e^{\left( \sigma ^{\prime }-\sigma \right) \gamma ^{\ast }-\left(
\sigma ^{\prime }-\sigma \right) ^{\ast }\gamma }  \notag \\
&=&\int \frac{\mathtt{d}^{2}\sigma ^{\prime }\mathtt{d}^{2}\sigma ^{\prime
\prime }}{4\pi ^{2}}\psi \left( \sigma ^{\prime }\right) \psi ^{\ast }\left(
\sigma ^{\prime \prime }\right) \int \mathtt{d}^{2}\gamma \delta \left( \eta
_{2}-D\frac{\sigma _{2}^{\prime }+\sigma _{2}^{\prime \prime }}{2}+B\gamma
_{1}\right)   \notag \\
&&\times \delta \left( \eta _{1}-D\frac{\sigma _{1}^{\prime }+\sigma
_{1}^{\prime \prime }}{2}-B\gamma _{2}\right) \exp \left\{ \allowbreak i%
\left[ \left( \sigma _{2}^{\prime }-\sigma _{2}^{\prime \prime }\right)
\gamma _{1}-\left( \sigma _{1}^{\prime }-\sigma _{1}^{\prime \prime }\right)
\gamma _{2}\right] \right\}   \notag \\
&=&\int \frac{\mathtt{d}^{2}\sigma ^{\prime }\mathtt{d}^{2}\sigma ^{\prime
\prime }}{4B^{2}\pi ^{2}}\psi \left( \sigma ^{\prime }\right) \psi ^{\ast
}\left( \sigma ^{\prime \prime }\right)   \notag \\
&&\times \exp \left\{ \allowbreak \frac{i}{B}\left[ \left( \sigma
_{2}^{\prime }-\sigma _{2}^{\prime \prime }\right) \left( -\eta _{2}+D\frac{%
\sigma _{2}^{\prime }+\sigma _{2}^{\prime \prime }}{2}\right) -\left( \sigma
_{1}^{\prime }-\sigma _{1}^{\prime \prime }\right) \left( \eta _{1}-D\frac{%
\sigma _{1}^{\prime }+\sigma _{1}^{\prime \prime }}{2}\right) \right]
\right\}   \notag \\
&=&\int \frac{\mathtt{d}^{2}\sigma ^{\prime }\mathtt{d}^{2}\sigma ^{\prime
\prime }}{4B^{2}\pi ^{2}}\psi \left( \sigma ^{\prime }\right) \psi ^{\ast
}\left( \sigma ^{\prime \prime }\right) \exp \left\{ \frac{i}{2B}\left[
D\left( \left\vert \sigma ^{\prime }\right\vert ^{2}-\left\vert \sigma
^{\prime \prime }\right\vert ^{2}\right) -2\eta _{1}\left( \sigma
_{1}^{\prime }-\allowbreak \sigma _{1}^{\prime \prime }\right) -2\eta
_{2}\left( \sigma _{2}^{\prime }-\sigma _{2}^{\prime \prime }\right) \right]
\right\} .  \label{20}
\end{eqnarray}%
On the other hand, when the beam $\Phi \left( \eta \right) $ propagates
through the $\left[ D\left( -B\right) \left( -C\right) A\right] $ optical
system, according to the Fresnel integration (\ref{10})-(\ref{11}), we have%
\begin{eqnarray}
\left\vert \Psi \left( \eta ^{\prime }\right) \right\vert ^{2} &=&\int \frac{%
\mathtt{d}^{2}\eta }{\pi }\mathcal{K}_{2}^{\left( D,-B,-C\right) }\left(
\eta ^{\prime },\eta \right) \Phi \left( \eta \right) \int \frac{\mathtt{d}%
^{2}\eta ^{\prime \prime }}{\pi }\mathcal{K}_{2}^{\ast \left( D,-B,-C\right)
}\left( \eta ^{\prime },\eta ^{\prime \prime }\right) \Phi ^{\ast }\left(
\eta ^{\prime \prime }\right)   \notag \\
&=&\frac{1}{4B^{2}}\int \frac{\mathtt{d}^{2}\eta }{\pi }\exp \left\{ \frac{%
\mathtt{i}}{2B}\left[ -D\left\vert \eta \right\vert ^{2}+\left( \eta \eta
^{\prime \ast }+\eta ^{\ast }\eta ^{\prime }\right) -A\left\vert \eta
^{\prime }\right\vert ^{2}\right] \right\} \Phi \left( \eta \right)   \notag
\\
&&\times \int \frac{\mathtt{d}^{2}\eta ^{\prime \prime }}{\pi }\exp \left\{
\frac{\mathtt{i}}{2B}\left[ D\left\vert \eta ^{\prime \prime }\right\vert
^{2}-\left( \eta ^{\prime \prime \ast }\eta ^{\prime }+\eta ^{\prime \prime
}\eta ^{\prime \ast }\right) +A\left\vert \eta ^{\prime }\right\vert ^{2}%
\right] \right\} \Phi ^{\ast }\left( \eta ^{\prime \prime }\right)   \notag
\\
&=&\frac{1}{4B^{2}\pi ^{2}}\int \frac{\mathtt{d}^{2}\eta }{\pi }\Phi \left(
\eta \right) \Phi ^{\ast }\left( \eta ^{\prime \prime }\right) \exp \left\{
\frac{\mathtt{i}}{2B}\left[ D\left( \left\vert \eta ^{\prime \prime
}\right\vert ^{2}-\left\vert \eta \right\vert ^{2}\right) -2\eta
_{1}^{\prime }\left( \eta _{1}^{\prime \prime }-\eta _{1}\right) -2\eta
_{2}^{\prime }\left( \eta _{2}^{\prime \prime }-\eta _{2}\right) \right]
\right\} ,  \label{21}
\end{eqnarray}%
which is the same as $R_{2}\left( \eta _{1},\eta _{2}\right) $ in (\ref{20})$%
.$ So combining (\ref{20}), (\ref{10})-(\ref{11}) and (\ref{21}) we reach
the conclusion

\begin{eqnarray}
&&\left\vert \frac{1}{-2\mathtt{i}B}\int \frac{\mathtt{d}^{2}\eta }{\pi }%
\exp \left\{ \frac{\mathtt{i}}{-2B}\left[ D\left\vert \eta \right\vert
^{2}-\left( \eta \eta ^{\prime \ast }+\eta ^{\ast }\eta ^{\prime }\right)
+A\left\vert \eta ^{\prime }\right\vert ^{2}\right] \right\} \Phi \left(
\eta \right) \right\vert ^{2}  \notag \\
&=&\pi \int \delta \left( \eta _{1}^{\prime }-D\sigma _{1}-B\gamma
_{2}\right) \delta \left( \eta _{2}^{\prime }-D\sigma _{2}+B\gamma
_{1}\right) W\left( \sigma ,\gamma \right) \mathtt{d}^{2}\sigma \mathtt{d}%
^{2}\gamma ,  \label{22}
\end{eqnarray}%
where $AD-BC=1$. The physical meaning of Eq. (\ref{22}) is: when an input
field propagates through an optical $\left[ D\left( -B\right) \left(
-C\right) A\right] $ system, the energy density of the output field is equal
to the Radon transform of the two-mode entangled Wigner function of the
input field. So far as our knowledge is concerned, this conclusion seems
new. Eq. (\ref{22}) is the relationship between the input amplitude and
output one in spatial-domain. Next we turn to the frequency domain.

If taking the matrix element of $F^{\left( t,k\right) }$ in the $\left\vert
\xi \right\rangle $ representation which is conjugate to $\left\vert \eta
\right\rangle $, where the overlap $\left\langle \eta \right\vert \left. \xi
\right\rangle $ is $\left\langle \eta \right\vert \left. \xi \right\rangle =%
\frac{1}{2}\exp [(\xi \eta ^{\ast }-\xi ^{\ast }\eta )/2],$ we obtain the
2-dimensional GFT in its `frequency domain', i.e.,
\begin{eqnarray}
&&\frac{1}{\pi }\left\langle \xi ^{\prime }\right\vert F^{\left( t,k\right)
}\left\vert \xi \right\rangle =\int \frac{\mathtt{d}^{2}\eta \mathtt{d}%
^{2}\eta ^{\prime }}{\pi ^{3}}\left\langle \xi ^{\prime }\right\vert \left.
\eta ^{\prime }\right\rangle \left\langle \eta ^{\prime }\right\vert
F^{\left( t,k\right) }\left\vert \eta \right\rangle \left\langle \eta
\right\vert \left. \xi \right\rangle   \notag \\
&=&\frac{1}{4}\int \frac{\mathtt{d}^{2}\eta \mathtt{d}^{2}\eta ^{\prime }}{%
\pi ^{2}}\exp \left( \frac{\xi ^{\prime \ast }\eta ^{\prime }-\xi ^{\prime
}\eta ^{\prime \ast }+\xi \eta ^{\ast }-\xi ^{\ast }\eta }{2}\right)
\mathcal{K}^{\left( t,k\right) }\left( \eta ^{\prime },\eta \right)   \notag
\\
&=&\frac{1}{2\mathtt{i}\left( -C\right) \pi }\exp \left[ \frac{\mathtt{i}}{%
2\left( -C\right) }\left( D\left\vert \xi \right\vert ^{2}+A\left\vert \xi
^{\prime }\right\vert ^{2}-\xi ^{\prime \ast }\xi -\xi ^{\prime }\xi ^{\ast
}\right) \right] \equiv \mathcal{K}_{2}^{N}\left( \xi ^{\prime },\xi \right)
,  \label{23}
\end{eqnarray}%
where the superscript $N$ of $\mathcal{K}_{2}^{N}$ means that this transform
kernel corresponds to the parameter matrix $N=\left[ D,-C,-B,A\right] .$
Thus if the $\left[ D,-C,-B,A\right] $ system is changed to $\tilde{N}=\left[
A,C,B,D\right] $ system, the GFT in its `frequency domain' is given by%
\begin{equation}
\Psi \left( \xi ^{\prime }\right) =\int \mathcal{K}_{2}^{\tilde{N}}\left(
\xi ^{\prime },\xi \right) \Phi \left( \xi \right) \mathtt{d}^{2}\xi ,
\label{24}
\end{equation}%
where $\mathcal{K}_{2}^{\tilde{N}}\left( \xi ^{\prime },\xi \right) $ is%
\begin{equation}
\mathcal{K}_{2}^{N}\left( \xi ^{\prime },\xi \right) =\frac{1}{2\mathtt{i}%
C\pi }\exp \left[ \frac{\mathtt{i}}{2C}\left( A\left\vert \xi \right\vert
^{2}+D\left\vert \xi ^{\prime }\right\vert ^{2}-\xi ^{\prime \ast }\xi -\xi
^{\prime }\xi ^{\ast }\right) \right] .  \label{25}
\end{equation}%
It then follows from Eqs.(\ref{24}) and (\ref{25}) that%
\begin{eqnarray}
\left\vert \Psi \left( \xi ^{\prime }\right) \right\vert ^{2} &=&\int
\mathcal{K}_{2}^{\tilde{N}}\left( \xi ^{\prime },\xi \right) \Phi \left( \xi
\right) \mathtt{d}^{2}\xi \int \mathcal{K}_{2}^{\ast \tilde{N}}\left( \xi
^{\prime },\xi ^{\prime \prime }\right) \Phi ^{\ast }\left( \xi ^{\prime
\prime }\right) \mathtt{d}^{2}\xi ^{\prime \prime }  \notag \\
&=&\frac{1}{4\pi ^{2}C^{2}}\int \mathtt{d}^{2}\xi \mathtt{d}^{2}\xi ^{\prime
\prime }\Phi \left( \xi \right) \Phi ^{\ast }\left( \xi ^{\prime \prime
}\right)   \notag \\
&&\times \exp \left\{ \frac{\mathtt{i}}{2C}\left[ A\left( \left\vert \xi
\right\vert ^{2}-\left\vert \xi ^{\prime \prime }\right\vert ^{2}\right)
+2\xi _{1}^{\prime }\left( \xi _{1}^{\prime \prime }-\xi _{1}\right) +2\xi
_{2}^{\prime }\left( \xi _{2}^{\prime \prime }-\xi _{2}\right) \right]
\right\} .  \label{26}
\end{eqnarray}%
On the other hand, in similar to (\ref{18}), we consider the integration
transform$,$%
\begin{equation}
R_{2}\left( \xi _{1},\xi _{2}\right) =\pi \int \delta \left( \xi
_{1}-A\sigma _{1}-C\gamma _{2}\right) \delta \left( \xi _{2}-A\sigma
_{2}+C\gamma _{1}\right) W\left( \sigma ,\gamma \right) \mathtt{d}^{2}\sigma
\mathtt{d}^{2}\gamma ,  \label{27}
\end{equation}%
$R_{2}\left( \xi _{1},\xi _{2}\right) $ is also a probability distribution
along an infinitely thin phase space strip denoted by the real parameters $%
A,C$. Substituting (\ref{19}) into (\ref{27}) yields
\begin{eqnarray}
R_{2}\left( \xi _{1},\xi _{2}\right)  &=&\int \frac{\mathtt{d}^{2}\sigma
^{\prime }\mathtt{d}^{2}\sigma ^{\prime \prime }}{\pi ^{2}}\psi \left(
\sigma ^{\prime }\right) \psi ^{\ast }\left( \sigma ^{\prime \prime }\right)
\int \delta ^{\left( 2\right) }\left( 2\sigma -\sigma ^{\prime }-\sigma
^{\prime \prime }\right) \mathtt{d}^{2}\sigma \mathtt{d}^{2}\gamma   \notag
\\
&&\times \delta \left( \xi _{2}-A\sigma _{2}+C\gamma _{1}\right) \delta
\left( \xi _{1}-A\sigma _{1}-C\gamma _{2}\right) e^{\left( \sigma ^{\prime
}-\sigma \right) \gamma ^{\ast }-\left( \sigma ^{\prime }-\sigma \right)
^{\ast }\gamma }  \notag \\
&=&\int \frac{\mathtt{d}^{2}\sigma ^{\prime }\mathtt{d}^{2}\sigma ^{\prime
\prime }}{4\pi ^{2}}\psi \left( \sigma ^{\prime }\right) \psi ^{\ast }\left(
\sigma ^{\prime \prime }\right) \int \mathtt{d}^{2}\gamma \delta \left( \xi
_{2}-A\frac{\sigma _{2}^{\prime }+\sigma _{2}^{\prime \prime }}{2}+C\gamma
_{1}\right)   \notag \\
&&\times \delta \left( \xi _{1}-A\frac{\sigma _{1}^{\prime }+\sigma
_{1}^{\prime \prime }}{2}-C\gamma _{2}\right) \exp \left[ \frac{\sigma
^{\prime }-\sigma ^{\prime \prime }}{2}\gamma ^{\ast }-\frac{\sigma ^{\prime
\ast }-\sigma ^{\prime \prime \ast }}{2}\gamma \right]   \notag \\
&=&\int \frac{\mathtt{d}^{2}\sigma ^{\prime }\mathtt{d}^{2}\sigma ^{\prime
\prime }}{4\pi ^{2}C^{2}}\psi \left( \sigma ^{\prime }\right) \psi ^{\ast
}\left( \sigma ^{\prime \prime }\right)   \notag \\
&&\times \exp \left\{ \frac{\mathtt{i}}{2C}\left[ A\left( \left\vert \sigma
^{\prime }\right\vert ^{2}-\left\vert \sigma ^{\prime \prime }\right\vert
^{2}\right) +2\xi _{1}\left( \sigma _{1}^{\prime \prime }-\sigma
_{1}^{\prime }\right) +2\xi _{2}\left( \sigma _{2}^{\prime \prime }-\sigma
_{2}^{\prime }\right) \right] \right\} ,  \label{28}
\end{eqnarray}%
which is the same as $\left\vert \Psi \left( \xi ^{\prime }\right)
\right\vert ^{2}$ in (\ref{26})$.$ So combining (\ref{28}), (\ref{24})-(\ref%
{25}), and (\ref{26}) we can draw the conclusion%
\begin{eqnarray}
&&\left\vert \frac{1}{2\mathtt{i}C\pi }\int \exp \left[ \frac{\mathtt{i}}{2C}%
\left( A\left\vert \xi \right\vert ^{2}+D\left\vert \xi ^{\prime
}\right\vert ^{2}-\xi ^{\prime \ast }\xi -\xi ^{\prime }\xi ^{\ast }\right) %
\right] \Phi \left( \xi \right) \mathtt{d}^{2}\xi \right\vert ^{2}  \notag \\
&=&\pi \int \delta \left( \xi _{1}-A\sigma _{1}-C\gamma _{2}\right) \delta
\left( \xi _{2}-A\sigma _{2}+C\gamma _{1}\right) W\left( \sigma ,\gamma
\right) \mathtt{d}^{2}\sigma \mathtt{d}^{2}\gamma .  \label{29}
\end{eqnarray}%
This is the relationship between the output amplitude and input one's
entangled Wigner function in `frequency domain'.

In sum, based on the correspondence between Collins diffraction formula
(optical Fresnel transform) and the transformation matrix element of a
three-parameters two-mode squeezing operator in the entangled state
representation, we have explored the relationship between output field
intensity determined by the Collins formula and the input field's
probability distribution along an infinitely thin phase space strip. The
entangled Wigner function is introduced for recapitulating the result.

Work supported by the National Natural Science Foundation of China (Grant
Nos 10775097 and 10874174).

\end{document}